\documentclass[twocolumn,description,aps]{revtex4-1}
\usepackage{graphicx,epstopdf,amsmath,amssymb,multirow,CJK,color,tabularx}
\usepackage[german,english]{babel}
\usepackage[colorlinks, linkcolor=blue,anchorcolor=blue,citecolor=blue,urlcolor=blue]{hyperref}

\begin{document}
\title{Topological properties of non-centrosymmetric superconductors \emph{T}Ir$_{2}$B$_{2}$ (\emph{T}=Nb, Ta)}

\author{Yan Gao$^{1}$}
\author{Jian-Feng Zhang$^{1}$}
\author{Shengyuan A. Yang$^{2,3}$}
\author{Kai Liu$^{1}$}\email{kliu@ruc.edu.cn}
\author{Zhong-Yi Lu$^{1}$}\email{zlu@ruc.edu.cn}

\affiliation{$^{1}$Department of Physics and Beijing Key Laboratory of Opto-electronic Functional Materials $\&$ Micro-nano Devices, Renmin University of China, Beijing 100872, China}
\affiliation{$^{2}$Research Laboratory for Quantum Materials, Singapore University of Technology and Design, Singapore 487372, Singapore}
\affiliation{$^{3}$Center for Quantum Transport and Thermal Energy Science, School of Physics and Technology, Nanjing Normal University, Nanjing 210023, China}

\date{\today}

\begin{abstract}
A recent experiment reported two new non-centrosymmetric superconductors NbIr$_{2}$B$_{2}$ and TaIr$_{2}$B$_{2}$ with respective superconducting transition temperatures of 7.2~K and 5.2~K and further suggested their superconductivity to be unconventional [K. G$\acute{o}$rnicka \emph{et al}., Adv. Funct. Mater. 2007960 (2020)]. Here, based on first-principles calculations and symmetry analysis, we propose that \emph{T}Ir$_{2}$B$_{2}$ (\emph{T}=Nb, Ta) are topological Weyl metals in the normal state. In the absence of spin-orbit coupling (SOC), we find that NbIr$_{2}$B$_{2}$ has 12 Weyl points, and TaIr$_{2}$B$_{2}$ has 4 Weyl points, i.e. the minimum number under time-reversal symmetry; meanwhile, both of them have a nodal net composed of three nodal lines. In the presence of SOC, a nodal loop on the mirror plane evolves into two hourglass Weyl rings, along with the Weyl points, which are dictated by the nonsymmorphic glide mirror symmetry. Besides the rings, NbIr$_{2}$B$_{2}$ and TaIr$_{2}$B$_{2}$ have 16 and 20 pairs of Weyl points, respectively. The surface Fermi arcs are explicitly demonstrated. On the (110) surface of TaIr$_{2}$B$_{2}$, we find extremely long surface Fermi arcs ($\sim$0.6~${\text{\AA}}^{-1}$) located 1.4~meV below the Fermi level, which should be readily probed in experiment. Combined with the intrinsic superconductivity and the nontrivial bulk Fermi surfaces, \emph{T}Ir$_{2}$B$_{2}$ may thus provide a very promising platform to explore the three-dimensional topological superconductivity.
\end{abstract}

\date{\today} \maketitle

\section{INTRODUCTION}\label{sec_introduction}

Topological semimetals (TSMs) have stimulated great interest in the fields of condensed matter physics and materials science in the last decade~\cite{1CKChiu,2AABurkov,3NArmitage,4HGao,5XWan}. Among the various types of TSMs, Weyl semimetals (WSMs) are the most prominent one and are of particular interest due to their exotic properties, such as topological surface Fermi arc states~\cite{5XWan}, chiral anomaly~\cite{6HBNielsen,7SonSpivak}, chiral Landau bands~\cite{6HBNielsen,8ZhaoYang}, and negative longitudinal magnetoresistance~\cite{9HuangWF,10GaoNiu}. A WSM is characterized by the presence of two-fold degenerate Weyl points formed by linear crossing of valence and conduction bands near the Fermi level~\cite{5XWan,11WengBernevig}. These Weyl points can be further divided into type-I and type-II Weyl points according to the slopes of the involved bands~\cite{12SoluyanovBernevig,13XuZhang}. A type-I point has up-right Weyl-cone-like dispersion along any direction, whereas for a type-II point, there exists a certain direction along which the cone is tipped over. (This classification may also be extended to nodal lines, with the definition of type-I, type-II, and hybrid lines~\cite{14LiYaoYang,15GaoCohen,16ZhangYang}. Each Weyl point carries a topological charge (Chern number) $C=\pm1$, and it can be regarded as source (+1) or sink (-1) of the Berry curvature field in momentum spaces. The unit topological charge endows an isolated Weyl point with topological robustness, namely, it does not require any symmetry protection (except for the lattice translational symmetry)~\cite{17SunChang,18FangNagaosa,19XuDaiFang}; instead, it needs to break either the time reversal and/or the space inversion symmetry. Weyl points can also exist at generic \emph{k} point in the Brillouin zone. Because of these reasons, the existing approaches such as the symmetry indicators cannot be used to directly detect the Weyl points~\cite{20KhalafPo,21SongZhang,22JKruthoff,23YQian}, and the search for realistic WSM materials remains a challenging task.

\begin{figure*}[!t]
	\centering
	\includegraphics[width=0.86\textwidth]{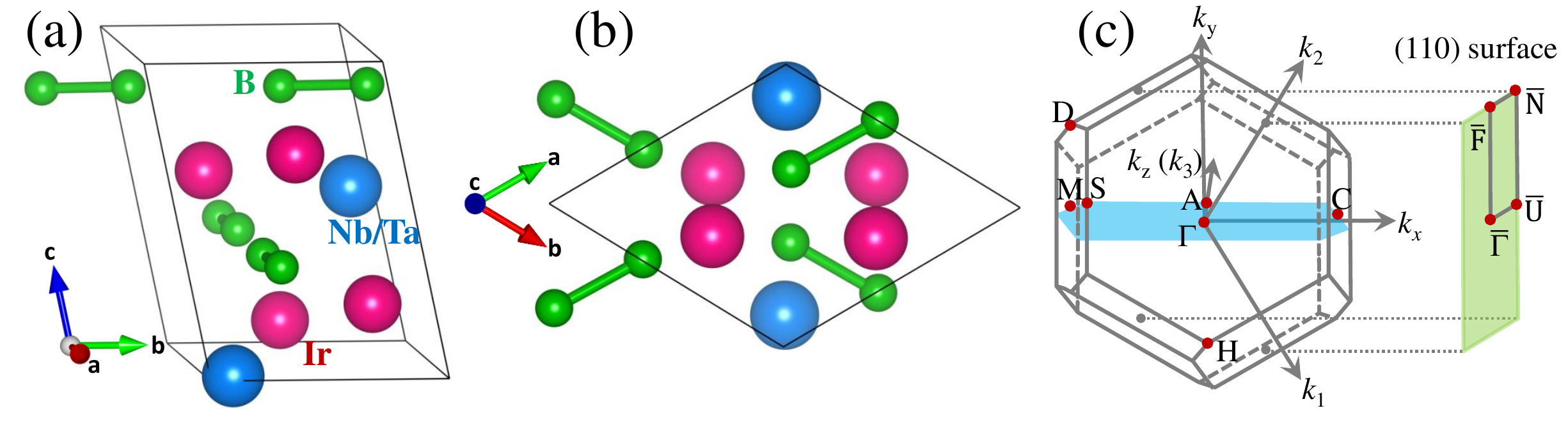}
	\caption{(Color online) (a) Side view and (b) top view of the primitive cell of \emph{T}Ir$_{2}$B$_{2}$ (\emph{T}=Nb, Ta). The blue, red, and green balls represent \emph{T}, Ir, and B atoms, respectively. (c) Bulk Brillouin zone (BZ) and projected 2D BZ of the (110) surface. The red dots and blue shaded plane represent the high-symmetry points and the glide mirror plane, respectively.}
	\label{fig_structure}
\end{figure*}

The coexistence of nontrivial topological properties and superconductivity in a material provides an opportunity to realize topological superconductivity and Majorana zero modes~\cite{24MSato,25SYGuan,26HJNoh,27KHJin,28JFZhang,29XuGu,30LAWray,31GaoGuo,32MKim,33QianNie}. Most of the current candidates for topological superconductivity are based on Fu-Kane's proposal~\cite{34FuKane}, that is, to achieve 2D topological superconductivity using the topological surface Dirac-cone states, e.g., it has been detected in recent experiments on iron-based superconductors~\cite{35DWang,36PZhang,37WLiu}. However, the study of three-dimensional (3D) topological superconductivity is limited due to the extreme lack of suitable materials~\cite{24MSato,38APMackenzie}. In this regard, it is worth noting that non-centrosymmetric WSMs with sign-changing superconductivity on Fermi surfaces carrying different Chern numbers can achieve 3D time-reversal symmetric topological superconductivity~\cite{39QiHughes,40PHosur}. Thus, it is highly desirable to search for materials in which the non-centrosymmetric WSM and the intrinsic superconductivity can coexist.

In a recent experiment, two new non-centrosymmetric materials NbIr$_{2}$B$_{2}$ and TaIr$_{2}$B$_{2}$ were synthesized and were found to be superconducting with $\emph{T}_c$ of 7.2~K and 5.2~K, respectively~\cite{41CavaKlimczuk}. Motivated by this experimental discovery, in this work, we investigate the electronic band structures of these two materials \emph{T}Ir$_{2}$B$_{2}$ (\emph{T}=Nb, Ta), based on first-principles calculations. We find that they both possess rich topological band features, including the Weyl points, the nodal net, the hourglass Weyl rings, and the topological surface Fermi arcs. Particularly, there exist extremely long ($\sim$0.6~${\text{\AA}}^{-1}$) Fermi arcs on the (110) surface close to the Fermi level, which should be readily detectable in experiment. Our work reveals the topological character of these newly synthesized materials. In view of their intrinsic superconductivity, our result also implies that these materials may provide a new class of material candidates for realizing 3D topological superconductivity.

\section{METHOD}\label{sec_method}

The electronic structures of \emph{T}Ir$_{2}$B$_{2}$ (\emph{T}=Nb, Ta) were studied with the density functional theory (DFT) calculations by using the projector augmented wave (PAW) method~\cite{42PAW} as implemented in the VASP package~\cite{43GKresse}. The generalized gradient approximation (GGA) of the Perdew-Burke-Ernzernof (PBE) type~\cite{44PBE} was adopted for the exchange-correlation functional. The kinetic energy cutoff of the plane wave basis was set to 500~eV. The Brillouin zone (BZ) of the primitive cell was sampled with an $8 \times 8 \times 6$ $\Gamma$-centered Monkhorst-Pack mesh~\cite{45HJMonkhorst}. The Gaussian smearing method with a width of 0.05~eV was adopted for the Fermi surface broadening. The energy and force convergence criteria were set to $10^{-6}$~eV and 0.01~eV/\text{\AA}, respectively. The optimized lattice parameters are $a=8.213~\text{\AA}$, $b=4.799~\text{\AA}$, $c=6.079~\text{\AA}$ for NbIr$_{2}$B$_{2}$ and $a=8.193~\text{\AA}$, $b=4.788~\text{\AA}$, $c=6.102~\text{\AA}$ for TaIr$_{2}$B$_{2}$, in good agreement with their experimental values ($a=8.159~\text{\AA}$, $b=4.775~\text{\AA}$, $c=6.007~\text{\AA}$ for NbIr$_{2}$B$_{2}$ and $a=8.133~\text{\AA}$, $b=4.763~\text{\AA}$, $c=6.021~\text{\AA}$ for TaIr$_{2}$B$_{2}$)~\cite{41CavaKlimczuk}. The topological charge (Chern number) for the Weyl points was computed by the Wilson-loop technique. The surface states and Fermi arcs of a semi-infinite \emph{T}Ir$_{2}$B$_{2}$ surface were calculated with the surface Green's function method by using the WannierTools~\cite{46QWu} package.

\section{RESULTS}\label{sec_results}

As shown in Figs.~\ref{fig_structure}(a) and~\ref{fig_structure}(b), the new ternary borides \emph{T}Ir$_{2}$B$_{2}$ (\emph{T}=Nb, Ta) reported in the recent experiment~\cite{41CavaKlimczuk} have the non-centrosymmetric space group No. 9 (\emph{Cc}) with the corresponding \emph{C}$_{s}$ point group symmetry. The group is nonsymmorphic and can be generated by the glide mirror $\widetilde{\mathcal{M}}_{y}$: $(x,y,z)\rightarrow(x,-y,z+\frac{1}{2})$. Moreover, the experiments have shown that \emph{T}Ir$_{2}$B$_{2}$ are nonmagnetic~\cite{41CavaKlimczuk}, which is also confirmed by our first-principles calculations, indicating that the time-reversal symmetry $\mathcal{T}$ is preserved. The corresponding bulk Brillouin zone (BZ) along with the high-symmetry \emph{k} points are displayed in Fig.~\ref{fig_structure}(c).

\begin{figure*}[!t]
	\centering
	\includegraphics[width=0.86\textwidth]{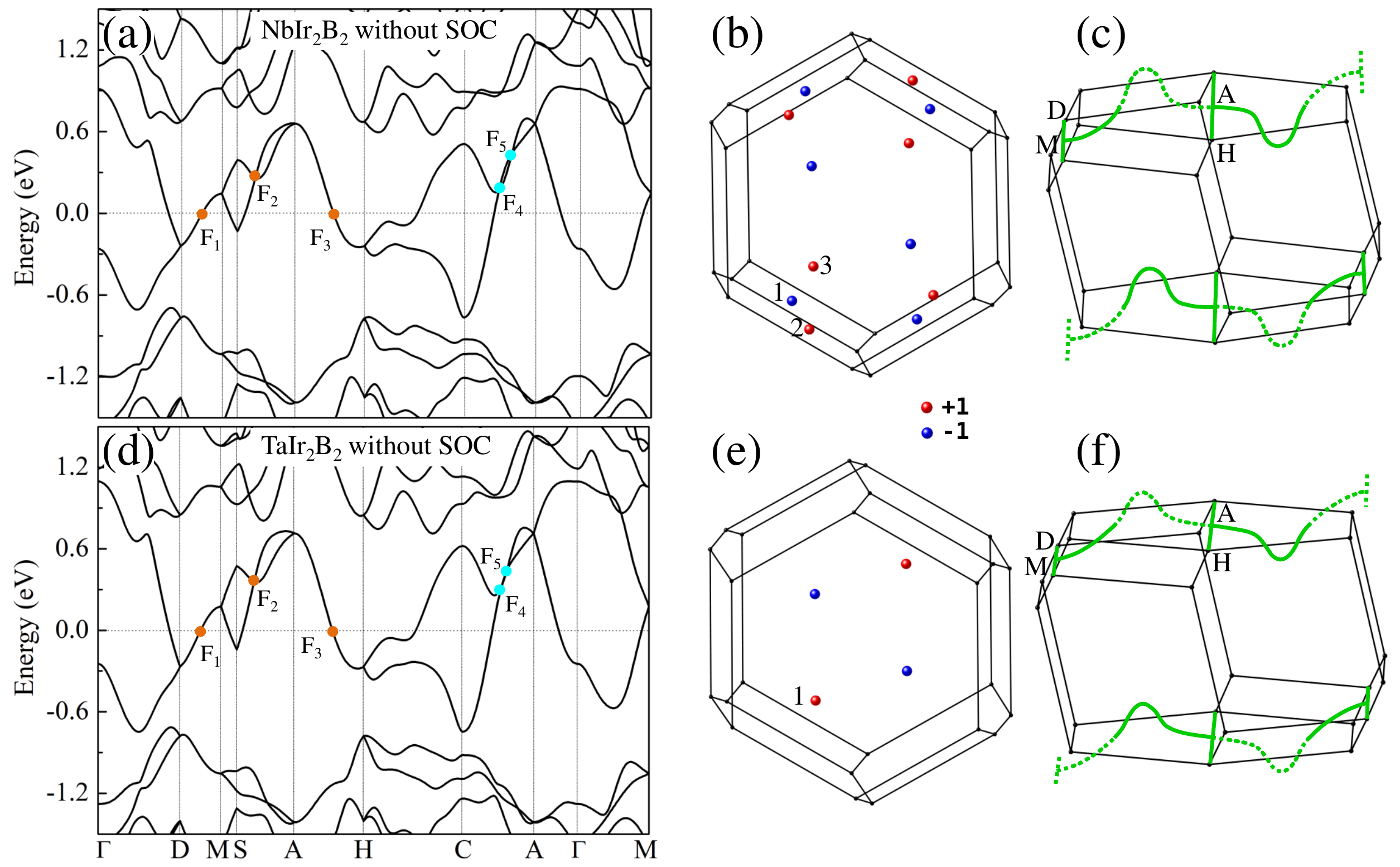}
	\caption{(Color online) Band structures of (a) NbIr$_{2}$B$_{2}$ and (d) TaIr$_{2}$B$_{2}$ calculated without the spin-orbit coupling (SOC). All Weyl points (red and blue dots) and the nodal net (green) of (b-c) NbIr$_{2}$B$_{2}$ and (e-f) TaIr$_{2}$B$_{2}$ without SOC. The red and blue dots represent the Weyl points with chiralities of +1 and -1, respectively. The green solid and dotted lines show the nodal lines within and outside the first BZ, respectively.}
	\label{fig_band}
\end{figure*}

The electronic band structures of NbIr$_{2}$B$_{2}$ and TaIr$_{2}$B$_{2}$ calculated in the absence of spin-orbit coupling (SOC) are shown in Figs.~\ref{fig_band}(a) and~\ref{fig_band}(d). One can see that NbIr$_{2}$B$_{2}$ and TaIr$_{2}$B$_{2}$ have very similar band structures, which can be understood as Nb and Ta are located in the same column in the periodic table and exhibit the same valence in these materials. There are two intersecting bands near the Fermi level, mainly consisting of Nb/Ta and Ir \emph{d} orbitals (see Fig. S1 in Supplemental Material (SM)~\cite{47SM}). The low-energy bands show several nontrivial characteristics. First, the two bands along the D-M and A-H paths form degenerate nodal lines (green) for the two materials. We have checked the band dispersions around these nodal lines. For example, the dispersions around points F$_{1}$ and F$_{3}$ of \emph{T}Ir$_{2}$B$_{2}$ along the \emph{k}$_{z}$ direction are shown in Fig. S2 in SM~\cite{47SM} with a type-I dispersion. Second, there are linear band-crossing point F$_{2}$ with opposite slope along the S-A path and points F$_{4}$ (F$_{5}$) with same slope along the C-A path (see Fig. S2 in SM~\cite{47SM}). A careful scan finds that these three points are not isolated but form another nodal loop connecting the two nodal lines on the D-M and A-H paths, which lead to a hybrid nodal loop (i.e., a loop consisting of both type-I and type-II points) in the $k_y=0$ plane. All these nodal lines form a nodal net in the entire BZ [Figs. \ref{fig_band}(c) and \ref{fig_band}(f)]. Furthermore, deviating from the high-symmetry paths, we find that there are respectively 12 and 4 isolated Weyl points in NbIr$_{2}$B$_{2}$ and TaIr$_{2}$B$_{2}$. The mirror and the time reversal operations connect four Weyl points, forming an equivalent group, so there are 3 and 1 nonequivalent Weyl points in NbIr$_{2}$B$_{2}$ [Fig.~\ref{fig_band}(b)] and TaIr$_{2}$B$_{2}$ [Fig.~\ref{fig_band}(e)], respectively. The existence of these Weyl points was further verified with the calculated chirality of each nonequivalent Weyl point by integrating the Berry curvature on a sphere enclosing it (see Fig. S3 in SM~\cite{47SM}). The positions and the energies of these nonequivalent Weyl points are listed in Table \ref{tab:I}.

\begin{table}[!b]
\caption{\label{tab:I} The Cartesian coordinates and energies of nonequivalent Weyl points of NbIr$_{2}$B$_{2}$ and TaIr$_{2}$B$_{2}$ calculated without SOC.}
\begin{center}
\begin{tabular*}{1.0\columnwidth}{@{\extracolsep{\fill}}cccc}
\hline\hline
 &   Weyl point  &  Position (${\text{\AA}}^{-1}$)  &  Energy (eV) \\
\hline
 \multirow{3}{*}{NbIr$_{2}$B$_{2}$} & WP$_{1}$ & (-0.3812, -0.5221, -0.0941) & -0.0644 \\
  & WP$_{2}$ & (-0.2802, -0.6698, -0.1578) & -0.1125 \\
  & WP$_{3}$ & (-0.2315, -0.2825, -0.3323) & 0.1572 \\
\hline
TaIr$_{2}$B$_{2}$ & WP$_{1}$ & (-0.2335, -0.3057, -0.3508) & 0.1473 \\
\hline\hline
\end{tabular*}
\end{center}
\end{table}

Here, we clarify the protection mechanisms of the nodal lines discussed above. First, for the hybrid node loop, it is located in the $k_y=0$ plane, which is the invariant subspace of the $\widetilde{\mathcal{M}_{y}}$ symmetry. Hence, each Bloch state $|\Phi_{n}\rangle$ in the plane can be selected as an eigenstate of $\widetilde{\mathcal{M}}_{y}$. One finds that
\begin{equation}\label{eq_1}
\begin{aligned}
  (\widetilde{\mathcal{M}}_{y})^{2} &= \mathcal{T}_{001} = e^{-ik_{z}},
\end{aligned}
\end{equation}
where $\mathcal{T}_{001}: (x,y,z)\rightarrow(x,y,z+c)$ is the lattice translation operator. Hence the eigenvalue $g_{y}$ of $\widetilde{\mathcal{M}}_{y}$ must be $\pm$$e^{-ik_{z}/2}$. Particularly, at the time-reversal invariant momentum (TRIM) points A: (0,0,$\pi$) and M: (-$\pi$,-$\pi$,$\pi$), we have $g_{y}=\pm$i, and
\begin{equation}\label{eq_2}
\begin{aligned}
  \widetilde{\mathcal{M}}_{y}(\mathcal{T}|\Phi_{n}\rangle) &= -g_{y}(\mathcal{T}|\Phi_{n}\rangle),
\end{aligned}
\end{equation}
where $\mathcal{T}$ is the time reversal operator. Hence even without the SOC, all the bands must have a double degeneracy at the A and M points, consisting of Kramers partner $|\Phi_{n}\rangle$ and $\mathcal{T}|\Phi_{n}\rangle$ with opposite eigenvalue $g_{y}$. This can be clearly seen in Figs.~\ref{fig_band}(a) and~\ref{fig_band}(d). And because of this degeneracy, the bands are paired up, and the low-energy band structure involves at least two bands. The two bands have opposite $\widetilde{\mathcal{M}}_{y}$ eigenvalues, so any possible crossing between them will form a nodal loop in the $\widetilde{\mathcal{M}}_{y}$ invariant mirror planes. This is the protection mechanism for the hybrid nodal loop in the $k_y=0$ plane.

\begin{figure*}[!th]
	\centering
	\includegraphics[width=0.88\textwidth]{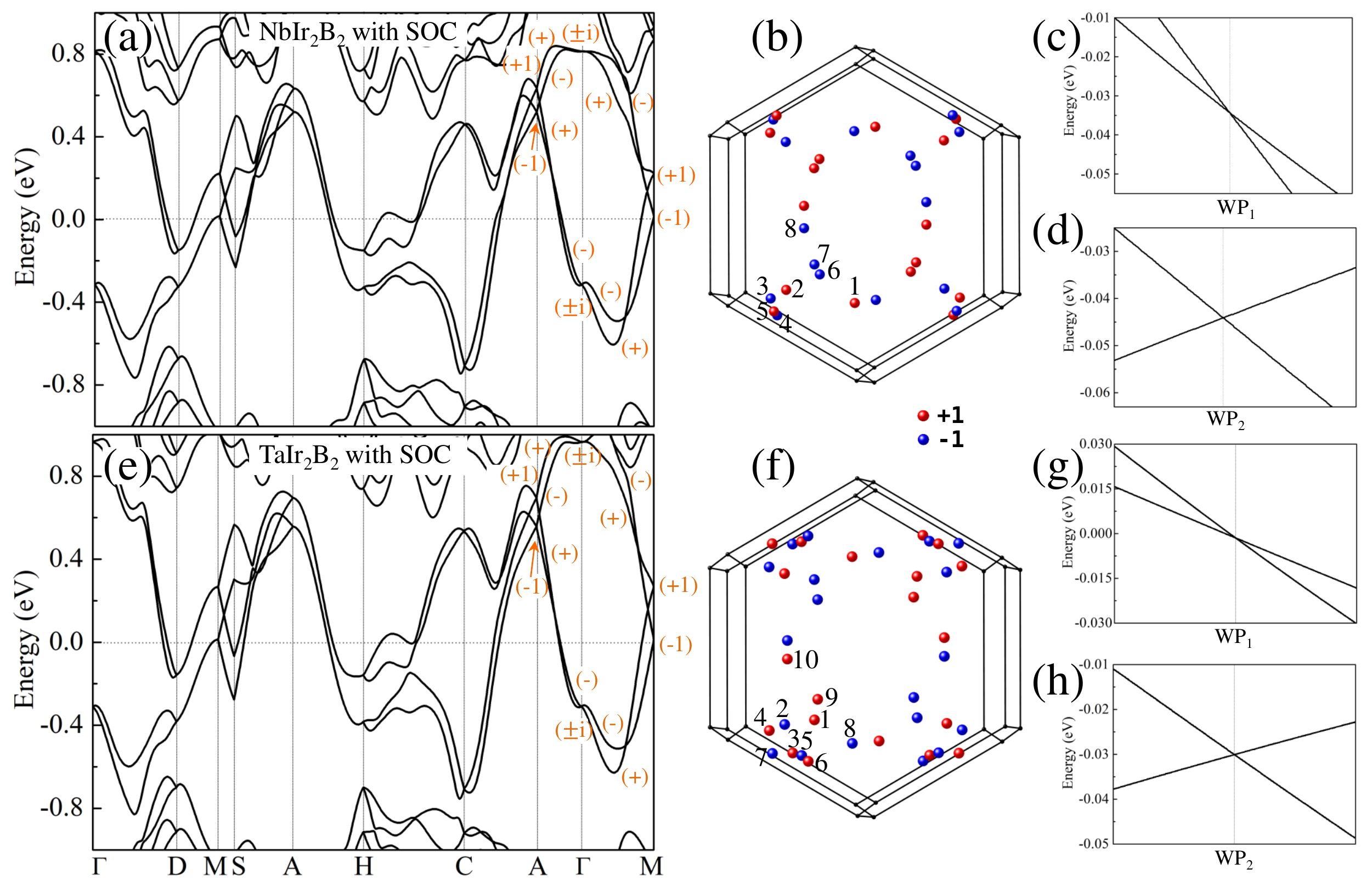}
	\caption{(Color online) Band structures of (a) NbIr$_{2}$B$_{2}$ and (e) TaIr$_{2}$B$_{2}$ calculated with SOC. The orange numbers and signs in the bands represent the eigenvalues $g_{y}$ of $\widetilde{\mathcal{M}}_{y}$. The perspective view of the Weyl points (red and blue) for (b) NbIr$_{2}$B$_{2}$ and (f) TaIr$_{2}$B$_{2}$. The red and blue dots represent the Weyl points with chiralities of +1 and -1, respectively. (c) and (d) display the type-II and type-I Weyl points along the \emph{k}$_{z}$ direction for WP$_{1}$ and WP$_{2}$ of NbIr$_{2}$B$_{2}$, respectively. (g) and (h) demonstrate the type-II and type-I Weyl points along the \emph{k}$_{z}$ direction for WP$_{1}$ and WP$_{2}$ of TaIr$_{2}$B$_{2}$, respectively.}
	\label{fig_SOCband}
\end{figure*}

Regarding the straight nodal line along paths D-M and A-H in the $k_z=\pi$ plane, the protection mechanism is different. On these high-symmetry paths, each \emph{k} point is invariant under the combined antiunitary symmetry $\widetilde{\mathcal{M}}_{y}\mathcal{T}$, and we have
\begin{equation}\label{eq_3}
\begin{aligned}
  (\widetilde{\mathcal{M}}_{y}\mathcal{T})^{2} &= e^{-ik_{z}} = -1.
\end{aligned}
\end{equation}
This means all the bands along the two paths must have a Kramer-like double degeneracy. In other words, the nodal lines along D-M and A-H paths are of essential band degeneracy features, which are guaranteed to exist by the nonsymmoprhic space group symmetry.

Next, we include the SOC effects. Then, the two nontrivial bands of \emph{T}Ir$_{2}$B$_{2}$ near the Fermi level split into four bands, as shown in Figs.~\ref{fig_SOCband}(a) and~\ref{fig_SOCband}(e). Among them, there are linear band-crossing points near the Fermi level along the M-S, M-$\Gamma$, A-S, and A-$\Gamma$ paths, which correspond to two nodal rings around the M and A point in the $k_y=0$ plane protected by the $\widetilde{\mathcal{M}}_{y}$ symmetry (see Fig. S4 in SM~\cite{47SM}). Actually, these are hourglass Weyl rings, around which the dispersion is of hourglass type, as can be seen from the band dispersion on the C-A, A-$\Gamma$, and $\Gamma$-M paths. The protection of this ring will be discussed in a while.

\begin{figure*}[!th]
	\centering
	\includegraphics[width=1.0\textwidth]{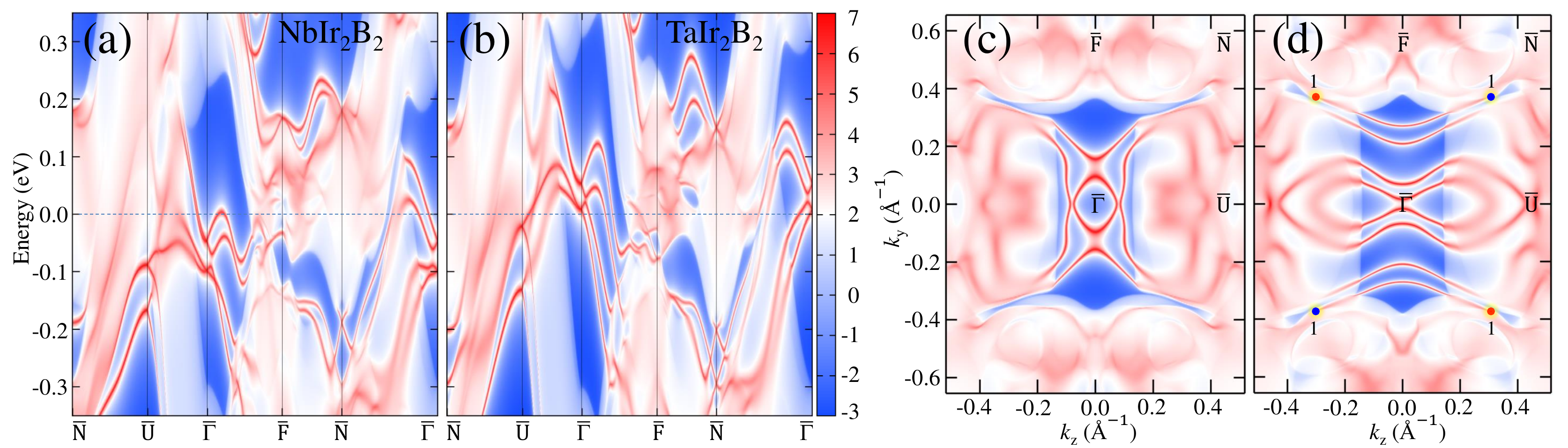}
	\caption{(Color online) Surface energy bands of (a) NbIr$_{2}$B$_{2}$ and (b) TaIr$_{2}$B$_{2}$ along the high-symmetry paths in the projected (110) surface. Surface spectra of (c) NbIr$_{2}$B$_{2}$ and (d) TaIr$_{2}$B$_{2}$ at a fixed energy E$_{f}$. Here the red and blue dots represent the four WP$_{1}$ points with chiralities of +1 and -1, respectively.}
	\label{fig_surfaces}
\end{figure*}

Besides the hourglass Weyl rings, the more important feature is the existence of Weyl points off the high-symmetry paths. We carefully scan the whole BZ and find that there are 32 and 40 Weyl points (red and blue dots) formed by the two low-energy bands for NbIr$_{2}$B$_{2}$ and TaIr$_{2}$B$_{2}$, respectively. Similar to the analysis for the case without the SOC, each Weyl point has another three partners related by the mirror and time-reversal symmetries, so in NbIr$_{2}$B$_{2}$ and TaIr$_{2}$B$_{2}$, there are 8 and 10 nonequivalent Weyl points in the entire BZ, respectively [Figs. \ref{fig_SOCband}(b) and \ref{fig_SOCband}(f)]. We have calculated the chirality of each nonequivalent Weyl points, as shown in Figs. S5 and S6 in the SM~\cite{47SM}. The positions and energies of these nonequivalent Weyl points are listed in Table \ref{tab:II}. These Weyl points are distributed in the energy interval from -0.11 to 0.39~eV around the Fermi level. Especially, the WP$_{1}$ and WP$_{2}$ of TaIr$_{2}$B$_{2}$ are about 1 and 30~meV below the Fermi level, and the WP$_{1}$ and WP$_{2}$ of NbIr$_{2}$B$_{2}$ about 34 and 44~meV below the Fermi level. They can be readily detected in ARPES experiment. We calculated the band dispersions around Weyl points WP$_{1}$ and WP$_{2}$ for NbIr$_{2}$B$_{2}$ [Figs.~\ref{fig_SOCband}(c) and~\ref{fig_SOCband}(d)] and TaIr$_{2}$B$_{2}$ [Figs.~\ref{fig_SOCband}(g) and~\ref{fig_SOCband}(h)]. One can observe that the Weyl cone is strongly tilted for WP$_{1}$ along the \emph{k}$_{z}$ direction, so that the electron- and hole-like pockets coexist at the same energy, hence it belongs to the type-II Weyl point. As for WP$_{2}$, it's Weyl cone is only slightly tilted and it belongs to the type-I Weyl point.

\begin{table}[!b]
\caption{\label{tab:II} The Cartesian coordinates and energies of all nonequivalent Weyl points of NbIr$_{2}$B$_{2}$ and TaIr$_{2}$B$_{2}$ calculated with SOC.}
\begin{center}
\begin{tabular*}{1.0\columnwidth}{@{\extracolsep{\fill}}cccc}
\hline\hline
 &   Weyl point  &  Position (${\text{\AA}}^{-1}$)  &  Energy (eV) \\
\hline
 \multirow{8}{*}{NbIr$_{2}$B$_{2}$} & WP$_{1}$ & (0.0132, -0.4606, 0.4944) & -0.0343 \\
  & WP$_{2}$ & (0.4101, -0.3954, 0.1539) & -0.0441 \\
  & WP$_{3}$ & (0.4994, -0.4424, 0.0803) & -0.0720 \\
  & WP$_{4}$ & (0.4697, -0.5334, 0.0208) & -0.0744 \\
  & WP$_{5}$ & (0.4857, -0.5124, 0.0404) & -0.0897 \\
  & WP$_{6}$ & (0.2097, -0.3088, 0.3898) & 0.1014 \\
  & WP$_{7}$ & (0.2470, -0.2574, 0.2874) & 0.1452 \\
  & WP$_{8}$ & (0.2862, -0.0599, 0.4710) & 0.3907 \\
\hline
 \multirow{10}{*}{TaIr$_{2}$B$_{2}$} & WP$_{1}$ & (-0.2348, -0.3732, -0.3971) & -0.0014 \\
  & WP$_{2}$ & (-0.4131, -0.4004, -0.1935) & -0.0301 \\
  & WP$_{3}$ & (-0.3837, -0.5538, -0.0495) & -0.0386 \\
  & WP$_{4}$ & (-0.5014, -0.4336, -0.1262) & -0.0417 \\
  & WP$_{5}$ & (-0.3342, -0.5668, -0.0689) & -0.0499 \\
  & WP$_{6}$ & (-0.2980, -0.5976, -0.0824) & -0.0701 \\
  & WP$_{7}$ & (-0.4902, -0.5559, -0.0673) & -0.0914 \\
  & WP$_{8}$ & (-0.0248, -0.4971, -0.4764) & -0.1095 \\
  & WP$_{9}$ & (-0.2244, -0.2650, -0.3238) & 0.1786 \\
  & WP$_{10}$ & (-0.3770, -0.0494, -0.4224) & 0.3136 \\
\hline\hline
\end{tabular*}
\end{center}
\end{table}

Now let's come back to the band structures in Figs.~\ref{fig_SOCband}(a) and~\ref{fig_SOCband}(e) and analyze several topological features. First of all, a twofold degenerate nodal line still exists along the D-M and A-H paths near the Fermi level. Second, most high-symmetry points (such as $\Gamma$, D, M, A, H, and C) have twofold band degeneracy. Third, along the A-$\Gamma$-M path, there are hourglass type dispersions, and the neck points of the hourglasses give twofold Weyl points on the hourglass Weyl rings.

To understand these features, we note that when the SOC is included, the glide mirror $\widetilde{\mathcal{M}}_{y}$ simultaneously acts on the spatial coordinates and the spin degrees of freedom, and we have
\begin{equation}\label{eq_4}
\begin{aligned}
  (\widetilde{\mathcal{M}}_{y})^{2} &= \mathcal{T}_{001}\bar{E} = -e^{-ik_{z}},
\end{aligned}
\end{equation}
where $\bar{E}$ represents the 2$\pi$ spin rotation. Because each \emph{k} point in the $k_y=0$ plane is invariant under the $\widetilde{\mathcal{M}}_{y}$, thus each Bloch state $|\Phi_{n}\rangle$ in the plane can be chosen as an eigenstate of $\widetilde{\mathcal{M}}_{y}$, and it's eigenvalue $g_{y}$ of $\widetilde{\mathcal{M}}_{y}$ must be $\pm$i$e^{-ik_{z}/2}$. Importantly, on the paths D-M and A-H, one finds that the previous Eq.~\ref{eq_3} for the combined symmetry $\widetilde{\mathcal{M}}_{y}\mathcal{T}$ remains valid even in the presence of SOC, as here we have $\mathcal{T}^{2}=-1$. Thus the Kramer-like degeneracy dictates a double degeneracy on these paths. This explains the first feature listed above. The second feature can also be easily understood. The points $\Gamma$, M, A, and C are TRIM points, so in the presence of SOC, the double degeneracy just corresponds to the Kramers degeneracy at these points. As for D and H points, they are points on the D-M and A-H paths, so the double degeneracy is from the $\widetilde{\mathcal{M}}_{y}\mathcal{T}$. Finally, regarding the third feature of the hourglass dispersion, we note that at the M point: (-$\pi$,-$\pi$,$\pi$) and the A point: (0,0,$\pi$), the eigenvalue $g_{y}=\pm1$ is real. Since M and A are TRIM points, which are invariant under $\mathcal{T}$, each Kramers pair $|\Phi_{n}\rangle$ and $\mathcal{T}|\Phi_{n}\rangle$ at A and M must have the same eigenvalue $g_{y}$. In contrast, at TRIM points $\Gamma$ and C, the eigenvalue $g_{y}=\pm$i is purely imaginary, so the Kramers pair there must have opposite eigenvalues. As a result, along an arbitrary path from M (A) to $\Gamma$ (C) on the mirror plane, there must be a partnering switching between the four band states, which must lead to an hourglass dispersion~\cite{48SigristSoluyanov,49WuJiao}. The neck point of the hourglass is a linear crossing between the bands of opposite $\widetilde{\mathcal{M}}_{y}$ eigenvalues, and it traces out a close ring on the mirror plane (see Fig. S4 in SM~\cite{47SM}). In Figs.~\ref{fig_SOCband}(a) and~\ref{fig_SOCband}(e), we have labeled the $g_{y}$ eigenvalues for the low energy states, which are consistent with our analysis.

\begin{figure*}[!t]
	\centering
	\includegraphics[width=0.92\textwidth]{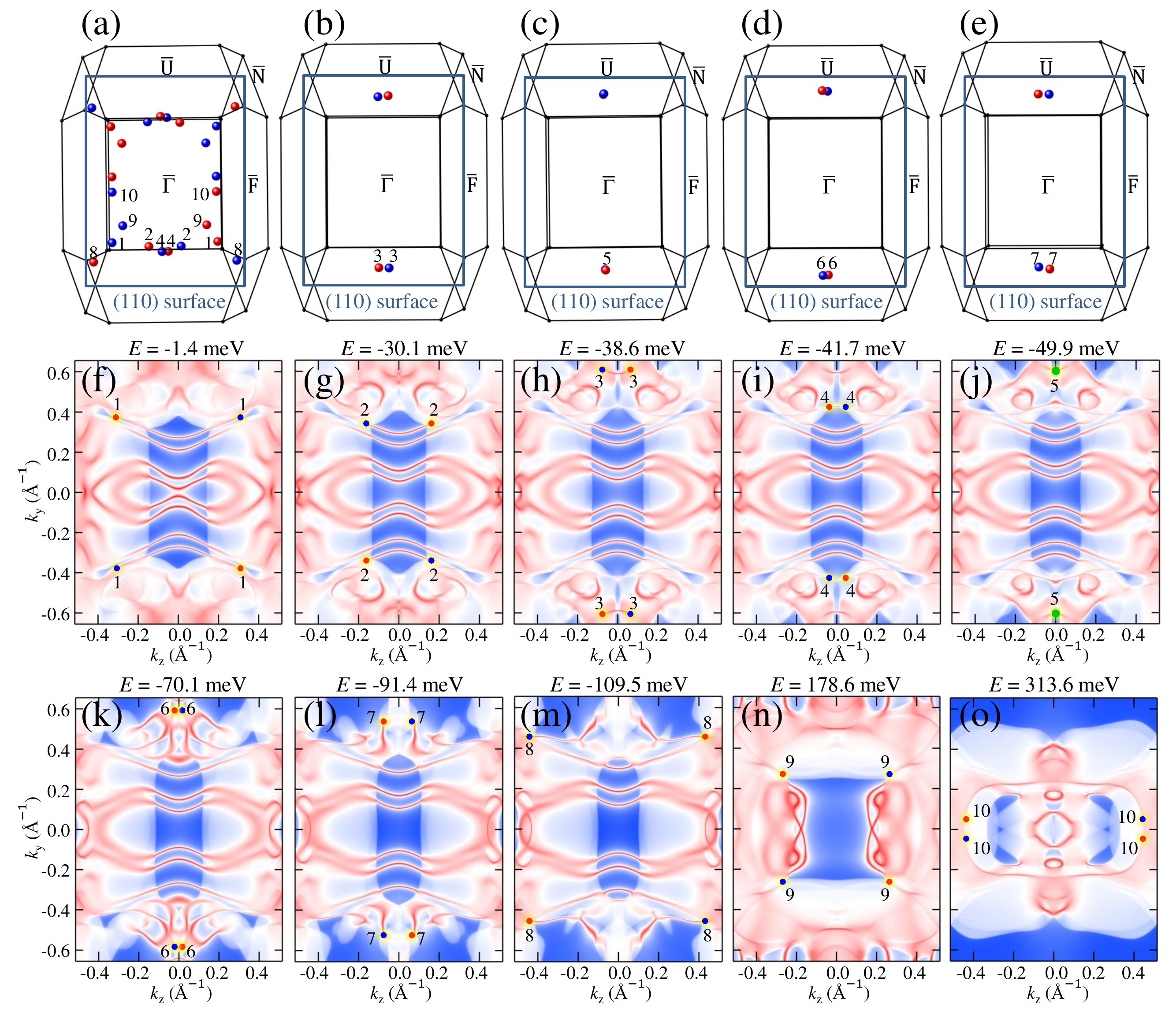}
	\caption{(Color online) (a-e) The projected twenty pairs of Weyl points of TaIr$_{2}$B$_{2}$ onto the (110) surface. Note that the mirror is projected in the $\bar{\Gamma}$-$\bar{F}$ direction. The Fermi surface of the projected (110) surface of TaIr$_{2}$B$_{2}$ with respect to the chemical potential is fixed at (f) $E=-1.4$~meV, (g) $E=-30.1$~meV, (h) $E=-38.6$~meV, (i) $E=-41.7$~meV, (j) $E=-49.9$~meV, (k) $E=-70.1$~meV, (l) $E=-91.4$~meV, (m) $E=-109.5$~meV, (n) $E=178.6$~meV, and (o) $E=313.6$~meV. The small red and blue dots represent the single projected Weyl points with chiralities of +1 and -1, respectively. The large green dots display two Weyl points with opposite chiralities projecting on top of each other as shown in (c) and (j).}
	\label{fig_Fermiarc}
\end{figure*}

A hallmark of the Weyl semimetal or metal is the existence of surface Fermi arcs connecting the projective Weyl points with opposite chiralities. The surface spectrum for the (110) surface of \emph{T}Ir$_{2}$B$_{2}$ are shown in Fig.~\ref{fig_surfaces}. As mentioned above, there are many nonequivalent Weyl points around the Fermi level, so the Fermi arcs form a rich pattern at different energy slices. Particularly, we can clearly find the Fermi arcs connecting the two projective WP$_{1}$ points for TaIr$_{2}$B$_{2}$ [Fig.~\ref{fig_surfaces}(d)], but it is difficult to find those for NbIr$_{2}$B$_{2}$ [Fig.~\ref{fig_surfaces}(c)]. This is because the WP$_{1}$ of the former is only 1.4~meV below the Fermi level. Moreover, we can see from Table \ref{tab:II} that TaIr$_{2}$B$_{2}$ has more Weyl points near the Fermi level. Therefore, we use TaIr$_{2}$B$_{2}$ as an example to clarify the patterns of these Fermi arcs.

In order to capture the connection pattern of the Fermi arc, by projecting the twenty pairs of Weyl points of TaIr$_{2}$B$_{2}$ onto the (110) surface [see Figs.~\ref{fig_Fermiarc}(a) to~\ref{fig_Fermiarc}(e)], we adjust the chemical potential to cross the 10 nonequivalent Weyl points (WP$_{1}$ to WP$_{10}$), corresponding to Figs.~\ref{fig_Fermiarc}(f) to~\ref{fig_Fermiarc}(o), respectively. We can clearly see an extremely long ($\sim$0.6~${\text{\AA}}^{-1}$) Fermi arc at 1.4~meV below the Fermi level that links a pair of the projective WP$_{1}$ with opposite chiralities along \emph{k}$_{z}$ [see Fig.~\ref{fig_Fermiarc}(f)]. By carefully checking the Fermi arcs, we find that the Fermi arcs are all connected with their own equivalent projective partners with opposite chiralities as shown in Figs.~\ref{fig_Fermiarc}(f) to~\ref{fig_Fermiarc}(o). Interestingly, a pair of WP$_{5}$ with opposite chiralities are projected onto the same position (large green dots) on the (110) surface [see Figs.~\ref{fig_Fermiarc}(c) and~\ref{fig_Fermiarc}(j)], which are interlinked by the Fermi arcs forming a closed loop around the $\bar{U}$ point.

\section{DISCUSSION AND SUMMARY}\label{sec_discussion}

Based on the above analysis, we have revealed the topological properties of the new ternary borides \emph{T}Ir$_{2}$B$_{2}$ (\emph{T}=Nb, Ta) synthesized in a recent experiment. Our calculations show that \emph{T}Ir$_{2}$B$_{2}$ not only has discrete Weyl points but also owns hourglass-type Weyl nodal rings protected by the nonsymmorphic symmetry, so both NbIr$_{2}$B$_{2}$ and TaIr$_{2}$B$_{2}$ are a kind of Weyl-point and Weyl-loop hybrid topological metals. Compared with other Weyl semimetal or metal materials, \emph{T}Ir$_{2}$B$_{2}$ has the following outstanding characters. First, the Weyl points of \emph{T}Ir$_{2}$B$_{2}$ distribute in a wide energy range from -0.11 to 0.39~eV, among which some are closer to the Fermi level than the case of TaAs family~\cite{11WengBernevig,50XuLee,51NeupaneZhang}. So even if there are a small amount of impurities and defects in the real samples, this will not affect the observation of their Weyl fermions. Second, the extremely long ($\sim$0.6~${\text{\AA}}^{-1}$) Fermi arcs of TaIr$_{2}$B$_{2}$ located 1.4~meV below the Fermi level are very rare in the reported Weyl semimetal or metal families, which will greatly promote the experimental detection of the Fermi arcs of \emph{T}Ir$_{2}$B$_{2}$. Third, combined with the intrinsic superconductivity and the nontrivial surface states derived from the Weyl points carrying the Chern number $C=\pm1$ near the Fermi level, \emph{T}Ir$_{2}$B$_{2}$ may serve as very promising candidates to explore the 2D and 3D topological superconductivity.

In summary, we have identified a class of non-centrosymmetric superconducting materials \emph{T}Ir$_{2}$B$_{2}$ (\emph{T}=Nb, Ta) as topological hybrid Weyl metals with Weyl points and hourglass Weyl rings. When the SOC is ignored, NbIr$_{2}$B$_{2}$ has 6 pairs of Weyl points and TaIr$_{2}$B$_{2}$ has the minimum number of four Weyl points under time-reversal symmetry, both of which possess a nodal net composed of three nodal lines. With the inclusion of SOC, a nodal loop on the mirror plane evolves into two hourglass Weyl rings enforced by the nonsymmorphic glide mirror symmetry; meanwhile, NbIr$_{2}$B$_{2}$ and TaIr$_{2}$B$_{2}$ have 16 and 20 pairs of Weyl points, respectively. The patterns of the Fermi arcs connecting the Weyl points with opposite chiralities are explicitly demonstrated, and the extremely long Fermi arcs ($\sim$0.6~${\text{\AA}}^{-1}$) near the Fermi level are also clearly observed on the (110) surface. Considering the nontrivial surface states near the Fermi level and the intrinsic superconductivity, \emph{T}Ir$_{2}$B$_{2}$ may offer a very promising material platform to realize the 2D and 3D topological superconductivity.

\begin{acknowledgments}

This work was supported by the National Key R\&D Program of China (Grants No. 2019YFA0308603 and No. 2017YFA0302903), the National Natural Science Foundation of China (Grants No. 11774422 and No. 11774424), the Beijing Natural Science Foundation (Grant No. Z200005), the CAS Interdisciplinary Innovation Team, the Fundamental Research Funds for the Central Universities, and the Research Funds of Renmin University of China (Grants No. 16XNLQ01 and No. 19XNLG13), and the Singapore Ministry of Education AcRF Tier 2 (MOE2017-T2-2-108). Y.G. was supported by the Outstanding Innovative Talents Cultivation Funded Programs 2021 of Renmin University of China. Computational resources were provided by the Physical Laboratory of High Performance Computing at Renmin University of China.

\end{acknowledgments}

\end{document}